\documentclass{article}
\usepackage{cite}
\usepackage{graphicx}
\usepackage{dcolumn}
\usepackage{fancyhdr}

\input{tcilatex}
\begin{document}

\date{}
\title{A most misunderstood conditionally-solvable quantum-mechanical model}
\author{Francisco M. Fern\'{a}ndez\thanks{%
fernande@quimica.unlp.edu.ar} \\
INIFTA, DQT, Sucursal 4, C. C. 16, \\
1900 La Plata, Argentina}
\maketitle

\begin{abstract}
In this paper we show that several authors have derived wrong physical
conclusions from a gross misunderstanding of the exact eigenvalues and
eigenfunctions of a conditionally-solvable quantum-mechanical model. It
consists of an eigenvalue equation with seemingly Coulomb, linear and
harmonic terms. Here we compare the results derived by those authors with
the actual eigenvalues of the models calculated by means of the Ritz
variational method.
\end{abstract}

\section{Introduction}

\label{sec:intro}

Several years ago Ver\c{c}in\cite{V91} obtained exact solutions to the
problem of two identical charged anyons moving in a plane under the
influence of a static uniform magnetic field perpendicular to that plane. He
derived an eigenvalue equation that is separable in cylindrical coordinates
so that the problem reduces to an eigenvalue equation for the radial part
with Coulomb plus harmonic terms. The application of the Frobenius method
leads to a three-term recurrence relation that the author used to truncate
the series in order to obtain exact polynomial solutions. From the results
obtained in this way the author concluded that ``there are bound states only
for certain discrete values of the magnetic field''. Later, Myrheim et al%
\cite{MHV92} (MHV from now on) discussed Ver\c{c}in's equations with more
detail finding that there are square-integrable solutions for all values of
the magnetic field. Therefore, the existence of allowed cyclotron
frequencies or allowed magnetic field intensities was proved to be an
artifact of the truncation method. Unfortunately, MHV did not stress this
point with sufficient clarity and left room for what we discuss in this
paper.

Independently, Taut discussed three quantum-mechanical models that also led
to a radial equation with Coulomb plus harmonic terms. They are two
electrons (interacting with Coulomb potentials) in an external
harmonic-oscillator potential\cite{T93}, two electrons (interacting with
Coulomb potentials) in a homogeneous magnetic field\cite{T94} and a
two-dimensional hydrogen atom in a homogeneous magnetic field\cite{T95}.
Taut followed the same mathematical procedure discussed above and made a
point that the truncation condition is sufficient but not necessary for
obtaining bound states. It is clear in these three papers that the
truncation of the three-term recurrence relation only yields some particular
states for some particular values of the oscillator frequency or
magnetic-field intensity.

Furtado et al\cite{FDMBB94} discussed the influence of a disclination on the
spectrum of an electron or a hole in a magnetic field in the framework of
the theory of defects. Although they were aware of the results derived by
both Ver\c{c}in\cite{V91} and MHV\cite{MHV92}, they surprisingly omitted the
latter more rigorous analysis and, based on the former, concluded that the
cyclotron frequency and the magnetic field should depend on the quantum
numbers. This mistake gave rise to a series of papers in which the authors
conjectured that cyclotron frequencies, oscillator frequencies, field
intensities and other physical quantities should have some particular
discrete values in order to have bound states\cite
{CB06,BM12,FB12,BB12,BB13,B14,B14b,BB14,FB15,BF15,OB16,VBB18,VFB18,VB18,BS18,VB19,BRS19,OMA19,LVB19,OBB20,VB20,VB20b,HMH20,A20,A20b,A21}
. The equations in these papers are separable in cylindrical coordinates
leading to an eigenvalue equation for the radial part with Coulomb plus
harmonic\cite
{V91,MHV92,FDMBB94,BM12,FB12,BB12,B14,B14b,BB14,FB15,BF15,OB16,VBB18,VB19,OMA19, VB20,VB20b,A21}%
, linear plus harmonic\cite
{CB06,FB12,B14b,VFB18,VB18,VB19,LVB19,OBB20,A20,A21} or Coulomb plus linear
plus harmonic terms\cite{CB06,FB12,BB13,BS18,VB19,BRS19,HMH20,A20b,A21}.

Several authors mentioned that the radial eigenvalue equation can be
transformed into the bi-confluent Heun equation\cite
{BM12,FB12,BB12,BB13,B14,B14b,BB14,FB15,BF15,OB16,VBB18,VFB18,VB18,
BS18,VB19 ,BRS19,OMA19,LVB19,OBB20,VB20,VB20b, HMH20,A21}; however, they did
not make use of any of the properties of the latter equation and simply
resorted to the straightforward Frobenius method.

In this paper we discuss the application of the Frobenius method to these
models and analyze the exact solutions obtained by truncation of the series
through the three-term recurrence relation. In section~\ref
{sec:general_model} we consider a general model that is separable in
cylindrical coordinates and leads to an eigenvalue equation for the radial
part that encompass the equations in all the papers mentioned above. In
section~\ref{sec:exact_sol} we apply the Frobenius method to the general
equation and show how to obtain some particular solutions in exact
analytical form. In sections \ref{sec:b=0} and \ref{sec:a=0} we show results
for two particular cases: Coulomb plus harmonic and linear plus harmonic
interactions, respectively. In section~\ref{sec:misinterpretation} we
discuss the misinterpretation of the exact results provided by the
truncation method mentioned above. Finally, in section~\ref{sec:conclusions}
we summarize the main results and draw conclusions.

\section{General model}

\label{sec:general_model}

In the papers listed above, the authors derived the eigenvalue equation for
the radial part of a wide variety of models with several different
interactions. In this section we introduce the main eigenvalue equation by
means of a simple, though quite general, quantum-mechanical model. It is
sufficient for present purposes to consider the Schr\"{o}dinger equation $%
H\psi =E\psi $ with the Hamiltonian operator
\begin{equation}
H=-\frac{\hbar ^{2}}{2m}\nabla ^{2}+V(\rho ),\;V(\rho )=\frac{V_{-2}}{\rho
^{2}}+\frac{V_{-1}}{\rho }+V_{1}\rho +\frac{m\omega ^{2}}{2}\rho ^{2},
\label{eq:H}
\end{equation}
where $m$ is the mass of the particle, $\rho =\sqrt{x^{2}+y^{2}}$, $V_{-2}>0$
and $V_{-1}$, $V_{1}$ real. Following a well known procedure for obtaining
dimensionless equations\cite{F20} we carry out the change of variables $%
\left( x,y,z\right) =\left( L\tilde{x},L\tilde{y},L\tilde{z}\right) $, where
$L=\sqrt{\hbar /(m\omega )}$, and define
\begin{eqnarray}
\tilde{H} &=&\frac{2}{\hbar \omega }H=-\tilde{\nabla}^{2}+\tilde{V}(\tilde{%
\rho}),  \nonumber \\
\tilde{V}(\tilde{\rho}) &=&\frac{2}{\hbar \omega }V(L\tilde{\rho})=\frac{%
2mV_{-2}}{\hbar ^{2}\tilde{\rho}^{2}}+\frac{a}{\tilde{\rho}}+b\tilde{\rho}+%
\tilde{\rho}^{2},  \nonumber \\
\tilde{\rho} &=&\frac{\rho }{L},\;a=\frac{2\sqrt{m}V_{-1}}{\hbar ^{3/2}\sqrt{%
\omega }},\;b=\frac{2V_{1}}{\sqrt{m\hbar }\omega ^{3/2}},  \label{eq:H_dim}
\end{eqnarray}
so that the dimensionless Schr\"{o}dinger equation becomes $\tilde{H}\tilde{%
\psi}=\tilde{E}\tilde{\psi}$, where $\tilde{E}=2E/(\hbar \omega )$.

In order to make the notation simpler, from now on we will omit the tilde on
the dimensionless coordinates. The Schr\"{o}dinger equation is separable in
spherical coordinates $\left( x=\rho \cos \phi ,\,y=\rho \sin \phi
,\,z=z\right) $ and we write the solution as $\psi (x,y,z)=e^{i(kz+l\phi
)}R(\rho )$, where $-\infty <k<\infty $ and $l=0\pm 1,\pm 2,\ldots $. In
this way we arrive at the following eigenvalue equation for the radial part
\begin{eqnarray}
&&\left[ \frac{1}{\rho }\frac{d}{d\rho }\rho \frac{d}{d\rho }-\frac{\gamma
^{2}}{\rho ^{2}}-\frac{a}{\rho }-b\rho -\rho ^{2}+W\right] R(\rho )=0,
\nonumber \\
&&\gamma ^{2}=\frac{2mV_{-2}}{\hbar ^{2}}+l^{2},\;W=\tilde{E}-k^{2}.
\label{eq:eig_eq}
\end{eqnarray}
We are interested in those solutions $R(\rho )$ that are square integrable:
\begin{equation}
\int_{0}^{\infty }\left| R(\rho )\right| ^{2}\rho \,d\rho <\infty ,
\label{eq:bound_states}
\end{equation}
which only take place for particular values of $W=W_{\nu ,\left| \gamma
\right| }(a,b)$, $\nu =0,1,\ldots $. It is convenient for present purposes
to label the eigenvalues with the value of $|\gamma |$ instead of the actual
quantum number $l$. Since the behaviour of $R(\rho )$ at origin and at
infinity is determined by the terms $\gamma ^{2}\rho ^{-2}$ and $\rho ^{2}$,
respectively, it is clear that there are square-integrable solutions for all
real values of $a$ and $b$. More precisely, the eigenvalues $W_{\nu ,\left|
\gamma \right| }(a,b)$ are continuous functions of $a$ and $b$ that satisfy
the Hellmann-Feynman theorem\cite{G32,F39}
\begin{equation}
\frac{\partial W}{\partial a}=\left\langle \frac{1}{\rho }\right\rangle >0,\;%
\frac{\partial W}{\partial b}=\left\langle \rho \right\rangle >0.
\label{eq:HFT}
\end{equation}
For this reason the allowed values of the energy of the system
\begin{equation}
E_{\nu ,\left| \gamma \right| }\left( a,b,\omega \right) =\frac{\hbar \omega
}{2}\left[ W_{\nu ,\left| \gamma \right| }\left( a,b\right) +k^{2}\right] ,
\label{eq:E_nu,gamma}
\end{equation}
are continuous functions of the model parameters $V_{-2}>0$, $-\infty
<V_{-1}<\infty $, $-\infty <V_{1}<\infty $ and $\omega >0$.

Slight variants of the eigenvalue equation (\ref{eq:eig_eq}) are well known
to be quasi-exactly solvable (or conditionally solvable) and have been
treated in several different ways; for example, by means of supersymmetric
quantum mechanics\cite{BCD17} (see also Turbiner's remarkable review\cite
{T16} and the references therein for other methods). In what follows we
discuss the approach followed in the papers mentioned above\cite
{V91,MHV92,FDMBB94,T93,T94,T95,CB06,BM12,FB12,BB12,
BB13,B14,B14b,BB14,FB15,BF15,OB16,VBB18,VFB18,BS18,VB19,BRS19,OMA19,OBB20,VB20,VB20b,HMH20,A21}%
.

\section{Exact solutions for the general case}

\label{sec:exact_sol}

Before proceeding, we want to make it clear that from now on we omit the
origin of the eigenvalue equation (\ref{eq:eig_eq}) and simply focus on its
solutions. In other words, we consider it to be the description of the
motion of a particle in a two-dimensional plane and will not take into
account the free motion along the $z$-axis. This strategy will facilitate
the comparison of present results with those in most of the papers cited
above. Note, for example, that the infinite degeneracy discussed by Ver\c{c}%
in\cite{V91} is not an issue here.

In order to obtain exact solutions to equation (\ref{eq:eig_eq}) we apply
the Frobenius method by means of the ansatz
\begin{equation}
R(\rho )=\rho ^{s}\exp \left( -\frac{b}{2}\rho -\frac{\rho ^{2}}{2}\right)
P(\rho ),\;P(\rho )=\sum_{j=0}^{\infty }c_{j}\rho ^{j},\;s=\left| \gamma
\right| .  \label{eq:ansatz}
\end{equation}
The expansion coefficients $c_{j}$ satisfy the three-term recurrence
relation
\begin{eqnarray}
c_{j+2} &=&A_{j}(a,b)c_{j+1}+B_{j}(W,b)c_{j},\;j=-1,0,1,2,\ldots
,\;c_{-1}=0,\;c_{0}=1,  \nonumber \\
A_{j}(a,b) &=&\frac{2a+b\left( 2j+2s+3\right) }{2\left( j+2\right) \left[
j+2\left( s+1\right) \right] },\;B_{j}(W,b)=\frac{4\left( 2j+2s-W+2\right)
-b^{2}}{4\left( j+2\right) \left[ j+2\left( s+1\right) \right] }.
\label{eq:TTRR}
\end{eqnarray}
If the truncation condition $c_{n+1}=c_{n+2}=0$, $c_{n}\neq 0$, $%
n=0,1,\ldots $, has physically acceptable solutions for $a$, $b$ and $W$
then we obtain exact eigenfunctions because $c_{j}=0$ for all $j>n$. This
truncation condition is equivalent to $B_{n}=0$, $c_{n+1}=0$ or
\begin{equation}
W_{s}^{(n)}=2\left( n+s+1\right) -\frac{b^{2}}{4},\;c_{n+1}(a,b)=0,
\label{eq:trunc_cond}
\end{equation}
where the second equation determines a relationship between the parameters $%
a $ and $b$. On setting $W=W_{s}^{(n)}$ the coefficient $B_{j}$ takes a
simpler form:
\begin{equation}
B_{j}\left( W_{s}^{(n)},b\right) =\frac{2\left( j-n\right) }{\left(
j+2\right) \left[ j+2\left( s+1\right) \right] }.  \label{eq:B_j_simpler}
\end{equation}
It is clear that the truncation condition (\ref{eq:trunc_cond}) cannot
provide all the bound-state solutions to the eigenvalue equation (\ref
{eq:eig_eq}) because it forces a relationship between the model parameters $%
a $ and $b$. As stated above there are bound states for all $-\infty
<a,b<\infty $ and those coming from the truncation condition are valid in a
considerably more restricted domain of these model parameters. More
precisely, there are bound states in the whole $a-b$ plane and polynomial
solutions only on some curves $c_{n+1}(a,b)=0$ in this plane. For this
reason, this kind of models is commonly called quasi-exactly solvable or
conditionally solvable\cite{BCD17,T16} (and references therein).

Since $B_{j}\left( W_{s}^{(n)},b\right) $ is independent of $a$ and $b$ and $%
A_{j}(-a,-b)=-A(a,b)$ we conclude that $c_{j}(-a,-b)=(-1)^{j}c_{j}(a,b)$.
The coefficient $c_{j}(a,b)$ is a polynomial function of order $j$ in each
of the variables $a$ and $b$; therefore, the condition $c_{n+1}(a,b)=0$ has
solutions of the form $a_{s}^{(n,i)}(b)$ or $b_{s}^{(n,i)}(a)$, $%
i=1,2,\ldots ,n+1$, and it can be proved that all the roots are real\cite
{CDW00,AF20}. The exact solutions to the radial eigenvalue equation (\ref
{eq:eig_eq}), given by the truncation method, are of the form
\begin{equation}
R_{s}^{(n,i)}(\rho )=\rho ^{s}\exp \left( -\frac{b}{2}\rho -\frac{\rho ^{2}}{%
2}\right) P^{(n,i)}(\rho ),\;P^{(n,i)}(\rho
)=\sum_{j=0}^{n}c_{j,s}^{(n,i)}\rho ^{j}.  \label{eq:R_s^(n,i)}
\end{equation}
These solutions already satisfy equations (\ref{eq:eig_eq}) and (\ref
{eq:bound_states}) but, as stated above, they are not the only allowed
solutions to the radial eigenvalue equation, a fact that is known since long
ago for the case $b=0$\cite{MHV92} (see also\cite{T93,T94,T95}).

For a given value of $b$ all the roots $W_{s}^{(n,i)}=W_{s}^{(n)}=2\left(
n+s+1\right) -\frac{b^{2}}{4}$, $i=1,2,\ldots ,n+1$, have the same value; on
the other hand, for a given value of $a$ the roots $W_{s}^{(n,i)}$, $%
i=1,2,\ldots ,n+1$, are points on the inverted parabola $W_{s}^{(n,i)}=2%
\left( n+s+1\right) -\frac{\left[ b_{s}^{(n,i)}\right] ^{2}}{4}$.

\section{First particular case $b=0$}

\label{sec:b=0}

This model has already been discussed in some of the papers listed above\cite
{V91,MHV92,FDMBB94,T93,T94,T95,BM12,FB12,BB12,B14,B14b,BB14,FB15,BF15,OB16,VBB18,VB19,OMA19,VB20,VB20b,A21}
and we analyze it in more detail in this section. In this case we have $%
W_{s}^{(n)}=2\left( n+s+1\right) $ and arrange the roots so that $%
a_{s}^{(n,i)}>a_{s}^{(n,i+1)}$, $i=1,2,\ldots ,n$. Since $%
c_{j}(-a)=(-1)^{j}c_{j}(a)$ then the roots of $c_{n+1}=0$ satisfy $%
a_{s}^{(n,i)}=-a_{s}^{(n,n+2-i)}$, $i=1,2,\ldots ,\frac{n+1}{2}$ for $n$ odd
and $a_{s}^{(n,i)}=-a_{s}^{(n,n+2-i)}$, $i=1,2,\ldots ,\frac{n}{2}$, $%
a_{s}^{(n,j)}=0$, $j=\frac{n}{2}+1$, for $n$ even. In other words, the roots
$a_{s}^{(n,i)}$ are symmetrically distributed with respect to the $W$ axis
in the $a-W$ plane. The authors of the papers just mentioned failed to
realize the existence of this multiplicity of roots\cite
{V91,FDMBB94,BM12,FB12,BB12,B14,B14b,BB14,FB15,BF15,OB16,VBB18,VB18,VB19,OMA19,LVB19,VB20,VB20b,A20,A21}%
.

It follows from the Hellmann-Feynman theorem (\ref{eq:HFT}) and the chosen
arrangement of roots that $\left( a_{s}^{(n,i)},W_{s}^{(n)}\right) $ is a
point on the curve $W_{i-1,s}(a)=W_{i-1,s}(a,0)$. In order to verify this
fact we need the actual eigenvalues $W_{\nu ,s}$ that can be obtained by
means of a suitable approximate method because the eigenvalue equation (\ref
{eq:eig_eq}) is not exactly solvable\cite{BCD17,AF20,T16}. Here, we resort
to the well known Rayleigh-Ritz variational method that is known to yield
upper bounds to all the eigenvalues\cite{P68} and, for simplicity, choose
the non-orthogonal basis set of Gaussian functions $\left\{ \varphi
_{j,s}(\rho )=\rho ^{s+j}\exp \left( -\frac{\rho ^{2}}{2}\right)
,\;j=0,1,\ldots \right\} $. It is worth noticing that the chosen basis set
takes into account the correct behaviour of the bound states at origin and
infinity. Besides, it is complete because the eigenfunctions of the
dimensionless two-dimensional harmonic oscillator with potential $V(\rho )=%
\frac{2mV_{-2}}{\hbar ^{2}\rho ^{2}}+\rho ^{2}$ are linear combinations of
these Gaussian functions. This basis set is far more practical than the one
discussed in an earlier discussion of the model\cite{AF20}. The actual
eigenvalues can also be obtained from the three-term recurrence relation (%
\ref{eq:TTRR}) as shown by MHV\cite{MHV92} but we find the Rayleigh-Ritz
method more straightforward. Figure~\ref{Fig:Wb0g0} shows the first
eigenvalues $W_{\nu ,0}(a)$ calculated in this way (blue, continuous lines)
and the roots $W_{0}^{(n)}$ given by the truncation condition (red points).
There is no doubt that the former connect the latter exactly as we stated
above. This figure makes it clear that the roots $W_{s}^{(n)}$ given by the
truncation condition are, by themselves, meaningless if one does not arrange
and connect them properly. Present curves $W_{\nu ,0}(a)$ are similar
(though for a different value of $s$) to those shown in Fig.~2 of MHV\cite
{MHV92} (their $\nu $ is straightforwardly related to present $W$).
Unfortunately, those authors did not show the positions of the exact
eigenvalues (given by the truncation method) on their continuous curves for
the actual eigenvalues of the model. Perhaps, it could have avoided the
misinterpretation that followed. Figure~\ref{Fig:Wb0g0} also shows that the
curves $W_{\nu ,0}(-a)$ (green, dashed lines) also connect the points $%
W_{0}^{(n)}$ in such a way that $\left( a_{0}^{(n,i)},W_{0}^{(n)}\right) $
is a point on the curve $W_{n+1-i,0}(-a)$. In general, the roots $%
W_{s}^{(n)} $ of the truncation condition are intersection points between
curves $W_{\nu ,s}(a)$ and $W_{\nu ^{\prime },s}(-a)$. This interesting fact
is due to the symmetry of the distribution of the roots obtained from the
truncation condition: $W_{s}^{(n)}(-a)=W_{s}^{(n)}(a)$. Another interesting
point is that the intersections between the curves $W_{\nu ,s}(a)$ and $%
W_{\nu ,s}(-a) $, $\nu =0,1,\ldots $, take place at $a=0$ and yield the
exact eigenvalues of the dimensionless two-dimensional harmonic oscillator
with potential $V(\rho )=\frac{2mV_{-2}}{\hbar ^{2}\rho ^{2}}+\rho ^{2}$, a
fact that is expected from what was said above. Figure~\ref{Fig:Wb0g0} also
shows an horizontal line at $W=W_{0}^{(10)}$ (red, dashed) that connects all
the roots $a_{0}^{(10,i)}$, $i=1,2,\ldots ,11$. Note that this figure shows
several aspects of the connection between the actual eigenvalues and the
roots of the truncation condition that were not addressed in earlier
discussions of the problem\cite{AF20,AF21}.

It should be clear, from the discussion above, that the exact eigenvalue $%
W_{s}^{(n)}$ is shared by $n+1$ different quantum-mechanical models given by
model parameters $a_{s}^{(n,i)}$, $i=1,2,\ldots ,n+1$. This fact is also
revealed, from a different angle, by the application of supersymmetric
quantum mechanics\cite{BCD17} and other suitable algebraic approaches\cite
{T16}.

\section{Second particular case $a=0$}

\label{sec:a=0}

Most of the features of this model, which has already appeared in some of
the papers listed above\cite
{CB06,FB12,B14b,VFB18,VB18,VB19,LVB19,OBB20,A20,A21}, are similar to those
of the preceding one. The reason is that $c_{j}(-b)=(-1)^{j}c_{j}(b)$,
where, for simplicity, we write $c_{j}(b)=c_{j}(0,b)$. Consequently, the
roots $b_{s}^{(n,i)}$ exhibit the same symmetry discussed above for $%
a_{s}^{(n,i)}$. The main difference is that in this case the roots $%
W_{s}^{(n,i)}=2\left( n+s+1\right) -\frac{\left[ b_{s}^{(n,i)}\right] ^{2}}{4%
}$ lie on an inverted parabola. We again arrange the roots as $%
b_{s}^{(n,i)}>b_{s}^{(n,i+1)}$, $i=1,2,\ldots ,n$, so that $\left(
b_{s}^{(n,i)},W_{s}^{(n,i)}\right) $ is a point on $%
W_{i-1,s}(b)=W_{i-1,s}(0,b)$. In order to obtain the actual eigenvalues $%
W_{\nu ,s}(b)=W_{\nu ,s}(0,b)$ we resort to the Rayleigh-Ritz variational
method and exactly the same basis set of Gaussian functions used in section~%
\ref{sec:b=0}.

Figure~\ref{Fig:Wa0g0} shows the roots $W_{0}^{(n,i)}$ given by the
truncation condition (red points) and the eigenvalues $W_{\nu ,0}(b)$ (blue,
continuous lines). We appreciate that the latter connect the former exactly
as argued above. This figure also shows that the eigenvalues $W_{\nu ,0}(-b)$
(green, dashed lines) also connect the roots $W_{0}^{(n,i)}$. As in the
preceding example, we conclude that $\left(
b_{s}^{(n,i)},W_{s}^{(n,i)}\right) $ is a point on $W_{n+1-i,s}(-b)$ and
that the roots $W_{s}^{(n,i)}$ appear at the intersections between lines $%
W_{\nu ,s}(b)$ and $W_{\nu ^{\prime },s}(-b)$. In particular, the
intersections between the curves $W_{\nu ,s}(b)$ and $W_{\nu ,s}(-b)$, $\nu
=0,1,\ldots $, take place at $b=0$ and yield the exact eigenvalues of the
two-dimensional harmonic oscillator with potential $V(\rho )=\frac{2mV_{-2}}{%
\hbar ^{2}\rho ^{2}}+\rho ^{2}$. Figure~\ref{Fig:Wa0g0} also shows the
inverted parabola that connects the points $\left(
b_{0}^{(15,i)},W_{0}^{(15,i)}\right) $, $i=1,2,\ldots ,16$ (red, dashed
line). Once again we want to stress that several of the features shown in
this figure were not addressed in an earlier discussion of this model\cite
{AF20,AF21}.

\section{Misinterpretation of the results}

\label{sec:misinterpretation}

For concreteness we restrict ourselves to the case $b=0$. The authors of
almost all the papers listed above were unaware of the multiplicity of roots
and just considered that the truncation condition $c_{n+1}(a)=0$ yields some
roots $a_{n,l}$, $n=1,2,\ldots $,\cite
{V91,FDMBB94,CB06,BM12,FB12,BB12,B14,B14b,BB14,FB15,BF15,OB16,VBB18,BS18,VB19,OMA19,VB20,VB20b,A21}%
. From them they derived analytical expressions for the frequencies
\begin{equation}
\omega _{n,l}=\frac{4mV_{-1}^{2}}{\hbar ^{3}a_{n,l}^{2}},
\label{eq:omega_n,l}
\end{equation}
and the energies
\begin{equation}
\mathcal{E}_{n,l}=\frac{\hbar \omega _{n,l}}{2}\left[ 2\left( n+s+1\right)
+k^{2}\right] .  \label{eq:E_n,l}
\end{equation}
They appeared to think that these are the actual energies of the
model and that there are allowed values of the frequency $\omega
_{n,l}$ that depend on the quantum numbers (in most of the cases
they considered $n$ to be a quantum number, which it is not, as
shown in the preceding sections). In other words, they conjectured
that there are no bound states except for some values of the
oscillator frequency (cyclotron frequency, magnetic-field
intensity, etc). This is the mistake first made by
Ver\c{c}in\cite{V91} and later corrected by MHV\cite{MHV92}. The
MHV paper is most relevant and if the authors mentioned above\cite
{FDMBB94,CB06,BM12,FB12,BB12,B14,B14b,BB14,FB15,BF15,OB16,VBB18,BS18,VB19,VB20,VB20b,A21}
had paid attention to it then they would not have misunderstood
the conditionally-solvable problem.

The true energies of this quantum-mechanical model are given by equation (%
\ref{eq:E_nu,gamma}) and are continuous functions of $a$, $b$ and $\omega $
(more precisely, of $V_{-2}$, $V_{-1}$, $V_{1}$ and $\omega $). This fact is
clearly shown in figure 2 of MHV\cite{MHV92} and in present figures \ref
{Fig:Wb0g0} and \ref{Fig:Wa0g0}.

\section{Conclusions}

\label{sec:conclusions}

\begin{figure}[tbp]
\begin{center}
\includegraphics[width=9cm]{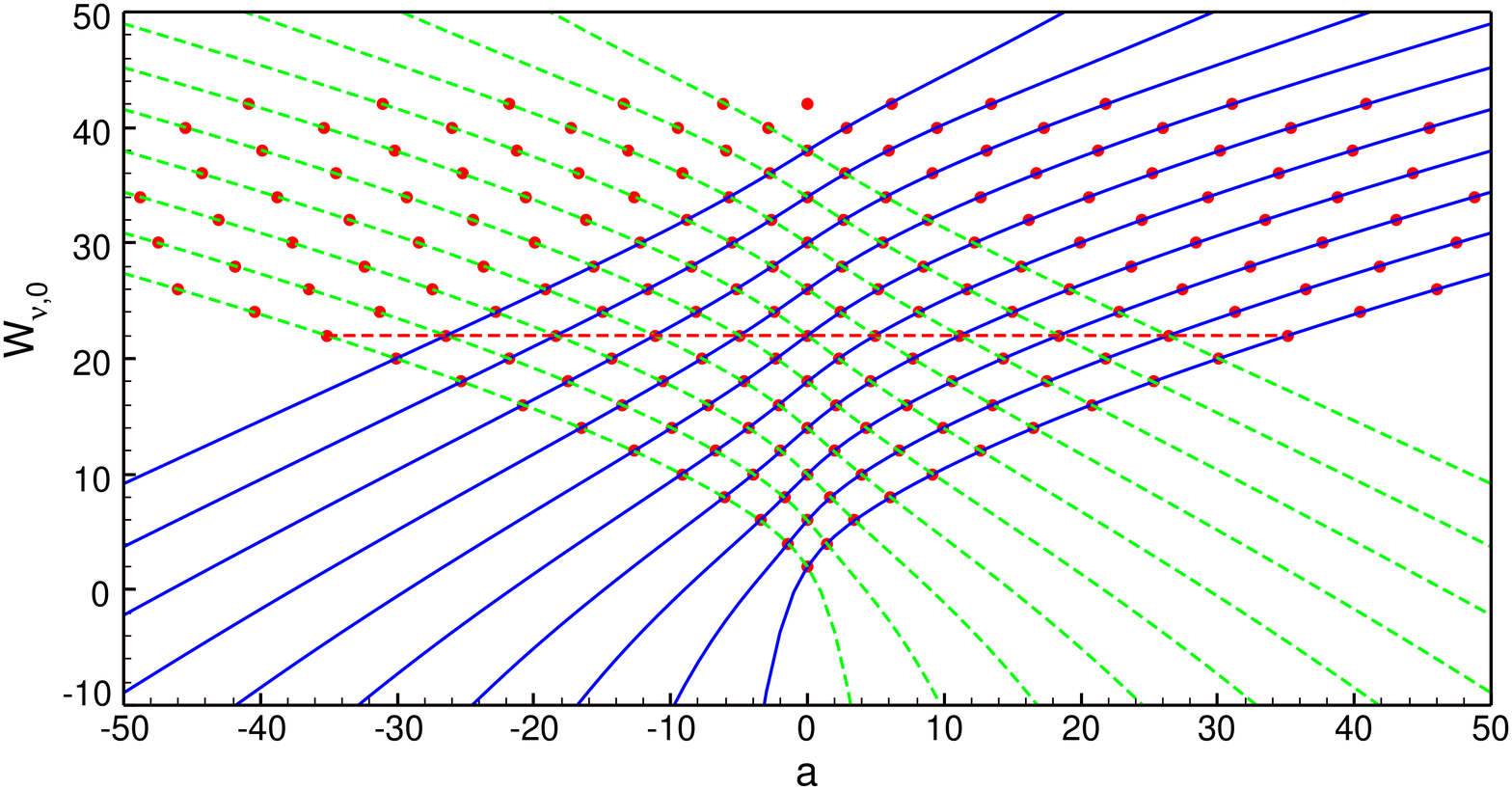}
\end{center}
\caption{Eigenvalues $W_0^{(n)}$ ($b=0$) from the truncation condition (red
points) and $W_{\nu,0}(a)$ (blue, continuous lines) and $W_{\nu,0}(-a)$
(green, dashed lines) obtained by means of the variational method}
\label{Fig:Wb0g0}
\end{figure}

\begin{figure}[tbp]
\begin{center}
\includegraphics[width=9cm]{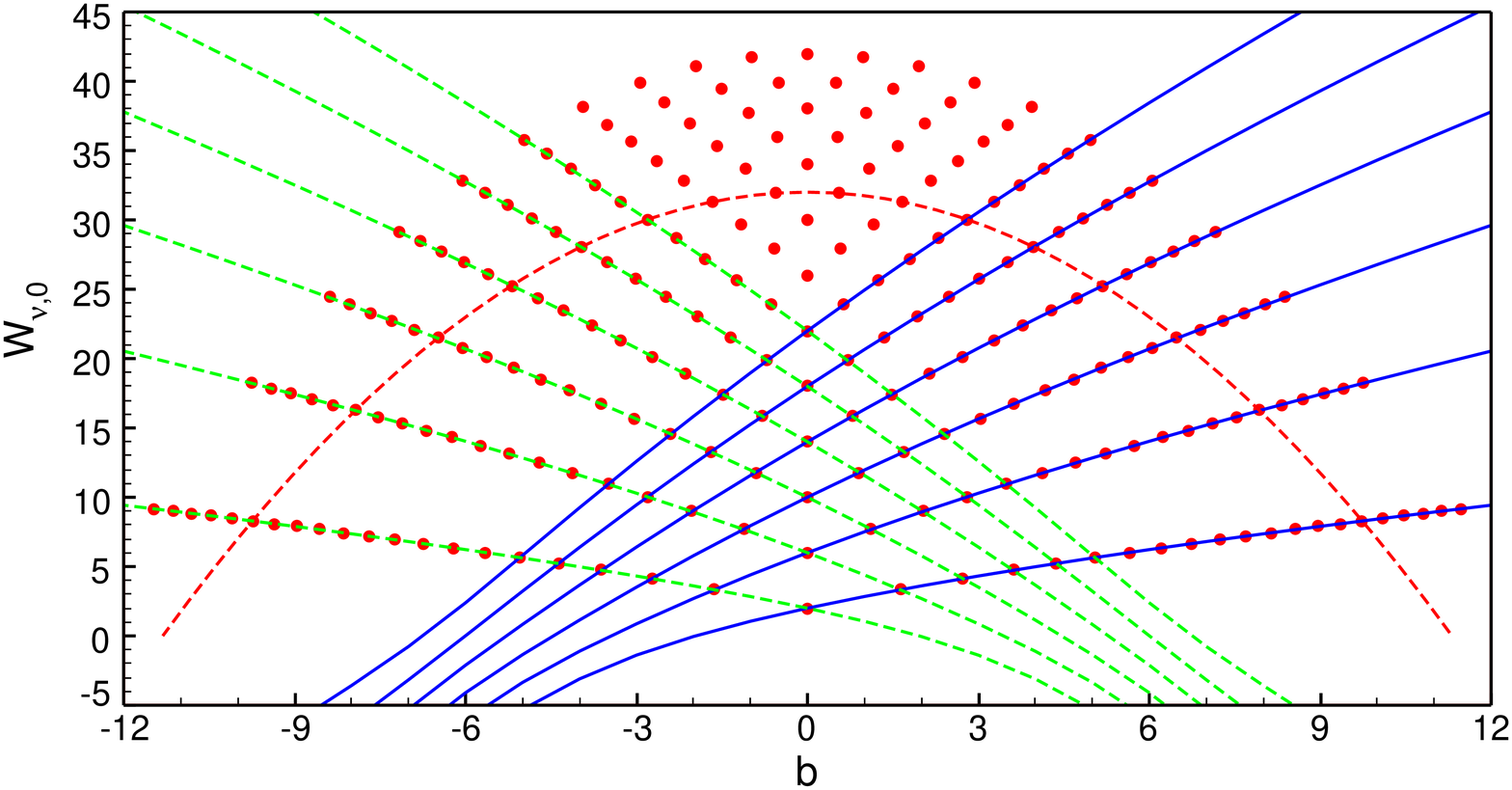}
\end{center}
\caption{Eigenvalues $W_0^{(n,i)}$ ($a=0$) from the truncation condition
(red points) and $W_{\nu,0}(b)$ (blue, continuous lines) and $W_{\nu,0}(-b)$
(green, dashed lines) obtained by means of the variational method}
\label{Fig:Wa0g0}
\end{figure}

Throughout this paper we have analyzed a series of papers\cite
{V91,MHV92,FDMBB94,T93,T94,T95,CB06,BM12,FB12,BB12,BB13,B14,B14b,BB14,FB15,BF15,OB16,VBB18,VFB18,VB18,BS18,VB19,BRS19,OMA19,LVB19,OBB20,VB20,VB20b, HMH20,A20,A20b,A21}
that exhibit the following features:

\begin{itemize}
\item  Their main equations are separable in cylindrical coordinates

\item  The eigenvalue equation for the radial part exhibits interactions
that resemble: Coulomb plus harmonic, linear plus harmonic or Coulomb plus
linear plus harmonic

\item  The authors applied the Frobenius method and derived three-term
recurrence relations for the expansion coefficients

\item  They obtained exact polynomial solutions by means of a simple
truncation condition

\item  Except in few cases\cite{MHV92,T93,T94,T95} the authors appeared to
believe that these are the only possible bound states

\item  They argued that those bound states occur only for particular values
of a cyclotron frequency, a magnetic-field intensity, or another model
parameter chosen for this purpose

\item  This mistake appears to stem from the well known fact that the
Frobenius method yields all the bound states for the harmonic oscillator,
the hydrogen atom and other exactly-solvable quantum-mechanical models were
the approach leads to a two-term recurrence relation\cite{P68} (this fact is
briefly addressed in the MHV paper\cite{MHV92}). Although the difference was
clearly discussed by MHV\cite{MHV92} their conclusions have been overlooked.
Consequently, those authors\cite
{MHV92,FDMBB94,T93,T94,T95,CB06,BM12,FB12,BB12,BB13,B14,B14b,BB14,FB15,BF15,OB16,VBB18,VFB18,VB18,BS18,VB19,BRS19,OMA19,LVB19,OBB20,VB20,VB20b, HMH20,A20,A20b,A21}
based their analysis on Ver\c{c}in's paper\cite{V91} and overlooked its
sequel were the mistake was corrected\cite{MHV92}

As shown above these models are conditionally solvable\cite{BCD17,CDW00,
AF20} and one obtains some particular bound states for some particular
relationships among model parameters. In this paper we have discussed the
connection between the eigenvalues given by the truncation condition, on the
one hand, and the true eigenvalues of the quantum-mechanical model, on the
other. Such relationship is made plain by figures \ref{Fig:Wb0g0} and \ref
{Fig:Wa0g0}. As argued above, present figures provide relevant information
omitted in earlier discussions of the problem\cite{AF20,AF21}. The roots $%
W_{s}^{(n.i)}$ by themselves are meaningless unless one is able to arrange
and connect them properly. The true eigenvalues are continuous functions of
the model parameters and, consequently, there are no allowed values of the
cyclotron frequency, magnetic-file intensity or the like. The existence of
polynomial solutions does not mean that allowed model parameters (model
parameters that depend on the quantum numbers) are physically meaningful
because the bound states are determined by equation (\ref{eq:bound_states})
and not by the truncation condition. Besides, as discussed in sections \ref
{sec:exact_sol}, \ref{sec:b=0} and \ref{sec:a=0}, the integer $n$ in the
truncation condition $c_{n+1}=0$ is by no means a quantum number as some of
the authors of the papers listed above appeared to believe.
\end{itemize}

\begin{thebibliography}{99}
\bibitem{V91}  Ver\c{c}in A 1991 \textit{Phys. Lett. B} \textbf{260} 120.

\bibitem{MHV92}  Myrheim J, Halvorsen E, and Ver\c{c}in A 1992 \textit{Phys.
Lett. B} \textbf{278} 171.

\bibitem{T93}  Taut M 1993 \textit{Phys. Rev. A} \textbf{48} 3561.

\bibitem{T94}  Taut M 1994 \textit{J. Phys. A} \textbf{27} 1045.

\bibitem{T95}  Taut M 1995 \textit{J. Phys. A} \textbf{28} 2081.

\bibitem{FDMBB94}  Furtado C, da Cunha B G C, Moraes F, Bezerra de Mello E
R, and Bezzerra V B 1994 \textit{Phys. Lett. A} \textbf{195} 90.

\bibitem{BM12}  Bakke K and Moraes F 2012 \textit{Phys. Lett. A} \textbf{376}
2838.

\bibitem{CB06}  A. L. Cavalcanti de Oliveira and E. R. Bezerra de Mello,
Exact solutions of the Klein-Gordon equation in the presence of a dyon,
magnetic flux and scalar potential in the spacetime of gravitational
defects, Class. Quantum Grav. 23 (2006) 5249-5263.

\bibitem{FB12}  Figueiredo Medeiros E R and Bezerra de Mello E R 2012
\textit{Eur. Phys. J. C} \textbf{72} 2051.

\bibitem{BB12}  Bakke K and Belich H 2012 \textit{Eur. Phys. J. Plus}
\textbf{127} 102.

\bibitem{BB13}  Bakke K and Belich H 2013 \textit{Ann. Phys. (Berlin)}
\textbf{526} 187.

\bibitem{B14}  Bakke K 2014 \textit{Ann. Phys.} \textbf{341} 86.

\bibitem{B14b}  Bakke K 2014 \textit{Int. J. Mod. Phys. A} \textbf{29}
1450117.

\bibitem{BB14}  Bakke K and Belich H 2014 \textit{Eur. Phys. J. Plus}
\textbf{129} 147.

\bibitem{FB15}  Fonseca I C and Bakke K 2015 \textit{J. Math. Phys.} \textbf{%
56} 062107.

\bibitem{BF15}  Bakke K and Furtado C 2015 \textit{Ann. Phys.} \textbf{355}
48.

\bibitem{OB16}  A. S. Oliveira and K. Bakke, Effcts on a Landau-type system
for a neutral particle with no permanent electric dipole moment subject to
the Kratzer potential in a rotating frame, Proc. Roy. Soc. A 472 (2016)
20150858.

\bibitem{VBB18}  Vit\'{o}ria L L, Bakke K, and Belich H 2018 \textit{Ann.
Phys.} \textbf{399} 117.

\bibitem{VFB18}  Vit\'{o}ria L L, Furtado C, and Bakke K 2018 \textit{Eur.
Phys. J. C} \textbf{78} 44.

\bibitem{VB18}  Vit\'{o}ria L L and Bakke K 2018 \textit{Int. J. Mod. Phys. D%
} \textbf{27} 1850005.

\bibitem{BS18}  Bakke K and Salvador C 2018 \textit{Proc. Roy. Soc. A}
\textbf{474} 20170881.

\bibitem{VB19}  Vit\'{o}ria L L and Belich H 2019 \textit{Phys. Scr.}
\textbf{94} 125301.

\bibitem{BRS19}  Bakke K, Ribeiro R F, and Salvador C 2019 \textit{Int. J.
Mod. Phys. A} \textbf{34} 1950229.

\bibitem{OMA19}  Oliveira A S, Maluf R V, and Almeida C A S 2019 \textit{%
Ann. Phys.} \textbf{400} 1.

\bibitem{LVB19}  Leite E V B, Vit\'{o}ria L L, and Belich H 2019 \textit{%
Mod. Phys. Lett. A} \textbf{34} 1950319.

\bibitem{VB20}  Vieira S L R and Bakke K 2020 \textit{Phys. Rev. A} \textbf{%
101} 032102.

\bibitem{VB20b}  Vit\'{o}ria L L and Belich H 2020 \textit{Eur. Phys. J. Plus%
} \textbf{135} 247.

\bibitem{OBB20}  Oliveira A S, Bakke K, and Belich H 2020 \textit{Eur. Phys.
J. Plus} \textbf{135} 623.

\bibitem{HMH20}  Hassanabadi H, de Montigny M, and Hosseinpour M 2020
\textit{Ann. Phys.} \textbf{412} 168040.

\bibitem{A20}  Ahmed F 2020 \textit{Eur. Phys. J. C} \textbf{80} 211.

\bibitem{A20b}  Ahmed F 2020 \textit{Eur. Phys. J. Plus} \textbf{135} 588.

\bibitem{A21}  Ahmed F 2021 \textit{Mod. Phys. Lett. A} \textbf{36} 2150004.

\bibitem{F20}  Fern\'{a}ndez F M 2020 Dimensionless equations in
non-relativistic quantum mechanics. arXiv:2005.05377 [quant-ph]

\bibitem{G32}  G\"{u}ttinger P 1932 \textit{Z. Phys.} \textbf{73} 169.

\bibitem{F39}  Feynman R P 1939 \textit{Phys. Rev.} \textbf{56} 340.

\bibitem{BCD17}  Bera S, Chakrabarti B, and Das T K 2017 \textit{Phys. Lett.
A} \textbf{381} 1356.

\bibitem{T16}  Turbiner A V 2016 \textit{Phys. Rep.} \textbf{642} 1.
arXiv:1603.02992 [quant-ph].

\bibitem{CDW00}  Child M S, Dong S-H, and Wang X-G 2000 \textit{J. Phys. A}
\textbf{33} 5653.

\bibitem{AF20}  Amore P and Fern\'{a}ndez F M 2020 \textit{Phys. Scr.}
\textbf{95} 105201. arXiv:2007.03448 [quant-ph]

\bibitem{P68}  Pilar F L 1968 \textit{Elementary Quantum Chemistry}
(McGraw-Hill, New York).

\bibitem{AF21}  P. Amore and F. M. Fern\'{a}ndez, An ubiquitous three-term
recurrence relation, J. Math. Phys. 62 (2021) 032106.
\end{thebibliography}
\end{document}